\documentclass[conference]{IEEEtran}
\IEEEoverridecommandlockouts
\usepackage{cite}
\usepackage{amsmath,amssymb,amsfonts}
\usepackage{graphicx}
\usepackage{textcomp}
\usepackage[T1]{fontenc}
\usepackage{xcolor}
\usepackage{booktabs}
\usepackage{multirow}
\usepackage{url}
\begin{document}
\title{Adversarial Robustness Is Not Zero-Day Generalization: A Controlled Evaluation of Flow-Based Intrusion Detectors on Corrected CIC-IDS2017}
\author{\IEEEauthorblockN{Abdulhamid Zaro}
\IEEEauthorblockA{\textit{College of Professions and Applied Sciences} \\
\textit{Hebron University}\\
Hebron, Palestine \\
abdulhamidz@hebron.edu \\
ORCID: 0009-0003-6241-302X}}
\maketitle
\begin{abstract}
Synthetic variants of known attacks, including feature-space adversarial perturbations, are sometimes used as stand-ins for unseen ("zero-day") attacks when evaluating machine-learning network intrusion detection systems (NIDS). We test whether this proxy holds. On the corrected CIC-IDS2017 release we train Random Forest, XGBoost, a 1D-CNN and an FT-Transformer and compare, at a fixed 1\% false-positive rate over five seeds, detection under constrained perturbations of known attacks (gradient-based and black-box attacks on 42 attacker-controllable features) with leave-one-attack-family-out (LOAO) detection, where an entire family is absent from training. Detection under the strongest perturbation we tested, with budgets of at most 0.2, stays between 90.9\% and 99.7\%, yet mean detection of held-out families falls to 76.2\% (RF), 40.2\% (XGBoost), 25.8\% (1D-CNN) and 22.8\% (FT-Transformer). Robustness exceeds LOAO detection for every model (gaps of 24 to 72 percentage points, 23 to 66 when robustness is also averaged over families; Holm-adjusted p $\leq$ 0.00096), and the two measures agree at most on a coarse ranking of the models. An exact mixed-integer evasion search confirms the tree ensembles' scores. Benign-only anomaly detectors and a simple multi-class open-set baseline do not close the gap. A temporal split on 5 seeds (3 of the 4 models) leaves the gap in place. On CSE-CIC-IDS2018 the gap recurs for every model (16 to 68 points). On these benchmarks and under this threat model, a robustness score should not be read as a zero-day score, and the two should be reported separately. Code and per-seed results are publicly available.
\end{abstract}

\section{Introduction}
Machine-learning NIDS routinely report near-perfect detection on public benchmarks, yet operational deployments face traffic and attacks that differ from the training distribution [1], [2]. The question that matters in operation is not how well a detector separates known attacks from benign traffic, but what it does with an attack it has never seen. Labelled zero-day traffic is scarce by definition, so recent work approximates it, either by evaluating across datasets [3] or by synthesising "new" attacks from known ones, for example with generative models [4], [5].

One synthesis route perturbs the feature vectors of known attacks with adversarial-example methods such as FGSM and PGD [6], [7], or interpolates between attack classes. If a detector withstands such perturbations, it is tempting to conclude that it will also generalise to novel attacks. That inference is rarely tested. A perturbed flow stays close to the attack it came from; a genuinely new attack family may sit somewhere else entirely.

We did not plan to write this paper. Our original design used perturbed and mixed attacks as a synthetic zero-day test set, with held-out attack families added only as a sanity check that the synthetic set behaved like real novelty. It failed that check. Detectors that barely moved under perturbation missed most flows of a family they had never seen, and the mismatch became the subject of the study.

We therefore test the inference directly. On the same data, models and operating point, we compare robustness to constrained perturbations of known attacks with detection of attack families that are absent from training. We make three contributions:
\begin{itemize}
\item A controlled protocol that evaluates the same four detectors under both regimes on the corrected CIC-IDS2017 release [8], [9], after removing duplicate and label-conflicting flows, at a fixed 1\% false-positive rate and over five seeds, with two unsupervised baselines as reference points.
\item A threat model that restricts perturbations to attacker-controllable features, projects them back to the feasible feature range, audits their internal consistency, and includes a score-based black-box attack [10] and a tree-aware coordinate search, so that tree ensembles are attacked directly rather than through gradients they do not have [11].
\item Evidence that perturbation robustness overstates unseen-family detection by 24 to 72 percentage points and tracks it only in coarse ranking, together with practical reporting recommendations.
\end{itemize}

\section{Related Work}
\subsection{Generalization and evaluation pitfalls in ML-based NIDS}
Sommer and Paxson explained early why machine learning struggles to find novel attacks in network traffic [1], and Arp et al. catalogue pitfalls that inflate results in security ML, including data snooping and inappropriate baselines [2]. Cross-dataset evaluation exposes large performance drops when NIDS meet traffic from another network [3]. For unknown attacks, a systematic review surveys zero-day detection approaches [12], and Zoppi et al. compare supervised, unsupervised and meta-learning detectors on attacks that are unknown at training time [13]. Several recent works instead generate "zero-day-type" samples from known attack data, for instance with a GAN trained on known malware records [4] or as synthetic attack signatures [5], and train or evaluate detectors on them, without testing whether such samples behave like genuinely new attack families. Generative models trained on known classes are likewise used to "synthesize samples of low-frequent and unknown attacks" [14] or traffic that "mimics zero-day patterns" [15], although the latter does evaluate with a leave-one-attack-type-out protocol. In adversarial-NIDS work, "zero-day" is sometimes applied to new evasion variants of known attacks, as when signatures of constrained attacks are proposed against "zero-day adversarial attacks" [16]; in this paper the term refers only to attack families absent from training. With a zero-shot protocol that holds out attack classes, Sarhan et al. found that, while most unseen classes are detected, detectors fail on certain attack groups whose features differ markedly from the training attacks [17], which is consistent with our LOAO results. The closest analogue outside security comes from computer vision, where Taori et al. found that robustness to synthetic image perturbations transfers little to natural distribution shifts across hundreds of models [18]; our study asks the same question for NIDS, with held-out attack families playing the role of the natural shift. Our LOAO protocol shares the goal of cross-dataset testing but holds the network environment fixed, isolating the novelty of the attack family, and it provides exactly the test that synthetic-proxy studies omit.

\subsection{Dataset quality of CIC-IDS2017}
CIC-IDS2017 [19] remains one of the most used NIDS benchmarks, but Engelen et al. documented systematic flow-construction and labelling errors and released corrected data [8]; follow-up work quantified their prevalence across datasets and released improved versions [9], and Lanvin et al. showed that such errors change reported detection performance substantially [20]. We use the corrected release and additionally remove exact duplicates and label-conflicting flows.

\subsection{Adversarial attacks on NIDS}
Gradient attacks [6], [7] require a differentiable model; tree ensembles need dedicated evasion methods [11] or black-box search [10]. Pierazzi et al. distinguish feature-space from problem-space attacks and formalise the constraints that realistic attacks must satisfy [21], and Apruzzese et al. model realistic adversarial attacks against NIDS [22]. Our perturbations are feature-space but constrained to attacker-controllable features; Section V-E measures how often they break flow-feature identities and repeats the experiment with those identities enforced.

\subsection{Position of this work}
Prior work typically measures either adversarial robustness or generalization to unknown attacks. To our knowledge, based on searches of OpenAlex and arXiv, no study measures both on identical models, data and operating point and tests whether one predicts the other; this is the gap we address.

\section{Methodology}
\subsection{Data and preprocessing}
We use the improved CIC-IDS2017 release of Liu et al. [9] (five capture days), which builds on the corrections of Engelen et al. [8]. Following the dataset authors' recommendation, flows labelled "Attempted" are relabelled benign. Identifier and leakage-prone fields (flow ID, source and destination IP and port, timestamp) are dropped, leaving 82 numeric flow features after removing constant columns. Rows with missing or infinite values are removed first, leaving 2,099,971 flows. Exact duplicates (identical features and label) are then removed, and feature vectors that occur under more than one label are discarded as ambiguous; together these two steps remove 384,774 flows, leaving 1,715,197. The label-conflict step uses labels from the whole dataset, but only to drop feature vectors that no detector could classify correctly under every label they carry, and it is applied identically before every split; Section VI gives its size. Attack labels are grouped into families: DoS, DDoS, PortScan, BruteForce (FTP/SSH-Patator), Botnet and WebAttack; Infiltration (36 flows) and Heartbleed (11 flows) are too small for separate evaluation and are excluded from training and testing. Family sizes after cleaning are given in Table~\ref{tab:data}.

\begin{table}[t]\centering\caption{Flows per family after cleaning, and per-seed train/test sizes used. Infiltration and Heartbleed are excluded from all experiments.}\label{tab:data}
\resizebox{\columnwidth}{!}{\begin{tabular}{lrrr}\toprule
Family & Unique flows & Train & Test \\\midrule
Benign & 1,432,904 & 80,000 (+20,000 val.) & 40,000 \\
DoS & 171,634 & 15,000 & 5,000 \\
DDoS & 95,144 & 15,000 & 5,000 \\
PortScan & 7,695 & 5,386 & 2,309 \\
BruteForce & 6,933 & 4,853 & 2,080 \\
Botnet & 736 & 515 & 221 \\
WebAttack & 104 & 72 & 32 \\
Infiltration (excluded) & 36 & -- & -- \\
Heartbleed (excluded) & 11 & -- & -- \\
\midrule Total & 1,715,197 & & \\
\bottomrule\end{tabular}}\end{table}

\subsection{Sampling and splits}
For each seed, every family is split 70/30 into disjoint train and test pools. From the pools we draw up to 100,000 benign training flows (20\% of which form a benign validation set used only to set thresholds), 40,000 benign test flows, and up to 15,000 training and 5,000 test flows per attack family. All scaling statistics are fitted on the training sample only and then applied unchanged to validation and test flows. The split is random at the flow level, so flows from the same attack session can fall on both sides; Section V-A discusses what this does to the in-distribution numbers.

\subsection{Detectors}
We evaluate Random Forest [23] (100 trees, balanced class weights), XGBoost [24] (300 trees, depth 8), a 1D-CNN (two convolutional layers with 32 and 64 filters, adaptive max-pooling and a 128-unit dense layer) and a compact FT-Transformer [25] (32-dimensional feature tokens, two encoder layers, four heads). The choice is deliberate. Tree ensembles remain the strongest and most widely used option on tabular flow features, while the two neural models stand for the convolutional and attention-based designs most often proposed for NIDS; having both families matters here because, as the results show, they fail in different ways. Neural models are trained for 10 epochs with Adam (learning rate $10^{-3}$) on signed-log, standardised features, with the positive class re-weighted for imbalance. All detectors are binary (benign vs. attack). The decision threshold of each model is set to the 99th percentile of its scores on the benign validation set (drawn from the training pool, never from test data), i.e. a target false-positive rate of 1\%; all detection rates are reported at this operating point.

As unsupervised reference points we add an Isolation Forest (200 trees) and a fully connected autoencoder (encoder 64-16, mirrored decoder, 20 epochs, reconstruction error as score), both trained on the benign training flows only and thresholded in the same way. They never see an attack, so every family is unseen for them; they answer the question of whether held-out families are detectable at all in this feature space at a 1\% alert budget. We did include a simple open-set baseline, a multi-class detector with maximum-softmax-probability rejection (Section V-C); dedicated open-set and meta-learning detectors [13] remain a natural extension.

\subsection{Threat model and constrained perturbations}
The attacker modifies flows of a known attack family to evade detection. Perturbations are applied only to the 42 attacker-controllable features: forward-direction packet counts, sizes and timing, flow duration, inter-arrival and active/idle statistics, and flow-level size and rate aggregates. The rule we followed was simple: an attacker controls what its own host sends and when, but not how the victim answers or how the protocol stack frames the exchange. Backward-direction features (server responses), TCP flags, protocol fields and initial window sizes are therefore left unchanged. Perturbations are bounded in $L_\infty$ norm by $\epsilon \in \{0.05, 0.1, 0.2, 0.5\}$ in the standardised signed-log feature space. In raw units, a step of $\epsilon$ multiplies $(1+|x|)$ of a modifiable feature by $\exp(\epsilon\,\sigma)$, where $\sigma$ is that feature's standard deviation in log space; across the 42 modifiable features the median factor is 1.18 at $\epsilon = 0.05$, 1.9 (interquartile range 1.4--2.8) at $\epsilon = 0.2$ and 5.0 (2.4--13) at $\epsilon = 0.5$. Budgets up to 0.2 thus amount to at most roughly doubling a typical feature, which an attacker controlling its own send rate and packet sizes can plausibly do; $\epsilon = 0.5$ is a large change that we keep as a stress test rather than as a realistic attacker, and it is not used in the robustness score that we compare with LOAO. After each perturbation, samples are projected back to a feasible region: values are clipped to the training range, integer-valued features are rounded, and every min/mean/max triplet is re-ordered so that min $\le$ mean $\le$ max.

We use six perturbation procedures: (i) a random-sign control with the same budget; (ii) FGSM [6] and (iii) 10-step PGD [7] computed on an independently trained MLP surrogate and transferred to every target; (iv) white-box PGD on the two neural targets; and (v) a score-based black-box coordinate search in the style of SimBA [10], which queries the target's score directly and therefore applies to tree ensembles. SimBA visits every modifiable feature in random order for two passes and keeps a $\pm\epsilon$ change whenever the attack score decreases; a five-pass variant is used as a convergence check (Section V-B). Because SimBA only tries the two corners of each coordinate's interval, it can step over the split points of a tree ensemble. We therefore add (vi) a tree-aware coordinate search for RF and XGBoost: for each modifiable feature it extracts the model's split thresholds that fall inside the flow's $\epsilon$-box, tries values just below and just above each of them (up to 24 per flow and feature) together with both box ends, and keeps the lowest-scoring one, for three passes. When at most 24 thresholds fall inside the box, this is exact one-dimensional minimisation of a piecewise-constant model along that coordinate; otherwise it searches a random subset of 24 of them.

\subsection{Cross-family hybrids}
As a second synthetic proxy, inspired by mixup [26], we form convex combinations $\lambda x_a + (1-\lambda) x_b$ of test flows from two different attack families, with $\lambda \sim U(0.3, 0.7)$, projected with the same feasibility step.

\subsection{Leave-one-attack-family-out (LOAO)}
For each family $f$, all detectors are retrained on benign traffic and the remaining five families, the threshold is re-set on the same benign validation set, and detection of the unseen family $f$ is measured. LOAO preserves the network environment and feature extraction and changes only whether the attack family was seen, making it a controlled proxy for zero-day attacks.

\subsection{Statistics}
Every configuration is run with five seeds (0 to 4), which change the splits, samples and model initialisation. We report means with 95\% Student-t confidence intervals across seeds. Changes in per-flow detection between clean and perturbed versions of the same flows are tested with an exact McNemar test [27], with Holm correction applied once across all 360 such tests (every model, procedure, budget and seed). We ask two separate questions about robustness and LOAO. The first is whether a robustness number overstates zero-day detection; both quantities are detection rates of attack flows at the same threshold, so we compare, per seed and model, the minimum detection rate over all attacks at $\epsilon \le 0.2$ (including the tree-aware search for the tree ensembles) with the mean LOAO detection rate. For brevity we call this minimum worst-case robustness; it is the worst case over the attacks we ran, not a certified bound, and a stronger attack could only lower it (Section VI). We also report a family-averaged variant that weights families as LOAO does. The comparison uses paired t-tests per model (Holm-adjusted across the four models, with the standardised effect size $d_z$) and a Wilcoxon signed-rank test over all model-seed pairs. The second is whether robustness predicts zero-day detection, i.e. whether more robust detectors also detect more unseen attacks; for this we use Spearman's rank correlation over the four model means, with an exact permutation test, because the model-seed pairs are not independent. Throughout, the primary metric is detection at the fixed 1\% false-positive rate. AUROC averages over all thresholds, including ones no operator would use, so it can stay high while detection at a realistic alert budget is low; we report it only as a secondary metric and return to the difference in Section V-F.

\section{Experimental Setup}
The main experiments ran on Google Colab with an NVIDIA Tesla T4 GPU, Python 3.13.16, PyTorch 2.11.0, XGBoost 3.4.1 and scikit-learn; the versions are those printed by the run itself and recorded in the accompanying log. One full run (five seeds, all scenarios) took 76 minutes. The open-set baseline and the temporal split were run later with the final version of the script on a CPU-only Colab runtime; its log records the library versions. All numbers in this paper are computed by script from the per-seed result files (the main file, results.csv, has 2,800 rows); none was entered by hand. The following are publicly available at https://github.com/abdulhamid-zaro/robustness-vs-zeroday-nids: the experiment script (all modes), the analysis script, the run logs with library versions, and nine result files: the main per-seed results, the anomaly-baseline and five-pass SimBA results, the tree-aware attack results, the identity audit, the repaired-perturbation results, the open-set results, the temporal-split results, the exact MILP results, and the CSE-CIC-IDS2018 results.

\section{Results and Discussion}
\subsection{Detection of known attacks}

On held-out flows of the six known families, every detector reaches near-ceiling performance at the 1\% operating point: detection rates of 100.0\% (RF), 100.0\% (XGBoost), 99.9\% (1D-CNN), 99.9\% (FT-Transformer), with F1 between 0.986 and 0.987 and measured false-positive rates within 0.1 points of the 1\% target (Table~\ref{tab:indist}; Table~\ref{tab:data} lists the data sizes). These numbers are a ceiling, not an estimate of field performance. The split is random at the flow level, so flows from the same attack session, which share almost identical statistics, appear on both sides; a temporal or session-level split would very likely lower them. None of our conclusions rests on them. They show only that the models are competent on what they have seen, which is the premise of the comparison that follows. To see how much of this ceiling comes from the split, we repeated the experiment with a temporal split (seeds 0, 1, 2, 3, 4): within every family, including benign traffic, the earliest 70\% of flows by start time form the training pool and the latest 30\% the test pool, so test flows come from later in each capture and the threshold is set on earlier benign traffic (Table~\ref{tab:temporal}). Known-attack detection becomes 99.5\% (RF), 99.8\% (XGBoost), 96.2\% (1D-CNN), and the false-positive rate measured on the later benign flows is 1.6\%, 1.4\%, 1.6\% against the 1\% target, so the threshold drifts on later traffic as one would expect.

\begin{table}[t]\centering\caption{Known attacks under the random flow-level split: detection rate, measured false-positive rate on benign test flows, F1 and AUROC at the threshold set for 1\% FPR on benign validation flows (mean $\pm$ 95\% CI over five seeds). These are ceilings, not field estimates (Section V-A).}\label{tab:indist}
\resizebox{\columnwidth}{!}{\begin{tabular}{lcccc}\toprule
Model & Detection & FPR & F1 & AUROC \\\midrule
RF & 1.0000 $\pm$ 0.0000 & 0.0098 $\pm$ 0.0013 & 0.987 $\pm$ 0.002 & 1.0000 \\
XGBoost & 0.9999 $\pm$ 0.0001 & 0.0103 $\pm$ 0.0013 & 0.986 $\pm$ 0.002 & 1.0000 \\
1D-CNN & 0.9995 $\pm$ 0.0003 & 0.0104 $\pm$ 0.0011 & 0.986 $\pm$ 0.001 & 0.9999 \\
FT-Transformer & 0.9991 $\pm$ 0.0005 & 0.0097 $\pm$ 0.0007 & 0.986 $\pm$ 0.001 & 0.9999 \\
\bottomrule\end{tabular}}\end{table}

\begin{table}[t]\centering\caption{Temporal split (earliest 70\% of each family for training, latest 30\% for testing; seed 0, 1, 2, 3, 4; FT-Transformer not run) against the random split on the same seed: known-attack detection and test FPR, worst-case robustness ($\epsilon \le 0.2$; FGSM, PGD, SimBA), mean LOAO detection and the gap (points).}\label{tab:temporal}
\resizebox{\columnwidth}{!}{\begin{tabular}{llccccc}\toprule
Model & Split & Known DR & Test FPR & Robustness & LOAO & Gap \\\midrule
RF & random & 1.000 & 0.010 & 0.998 & 0.762 & 23.6 \\
 & temporal & 0.995 & 0.016 & 0.985 & 0.696 & 28.9 \\
XGBoost & random & 1.000 & 0.010 & 0.998 & 0.402 & 59.5 \\
 & temporal & 0.998 & 0.014 & 0.982 & 0.372 & 61.0 \\
1D-CNN & random & 0.999 & 0.010 & 0.909 & 0.258 & 65.0 \\
 & temporal & 0.962 & 0.016 & 0.803 & 0.182 & 62.1 \\
\bottomrule\end{tabular}}\end{table}

\subsection{Robustness to constrained perturbations}

Table~\ref{tab:pert} and the per-seed results show that the tree ensembles change little under every perturbation procedure we tried: under the black-box SimBA attack, which queries the target directly, RF and XGBoost still detect 99.8\% and 99.3\% of perturbed attack flows even at $\epsilon = 0.5$, and none of the RF McNemar tests is significant after Holm correction at any budget. Running SimBA for five passes instead of two (200 flows per family and seed) leaves this unchanged: 99.7\% for RF and 99.2\% for XGBoost. SimBA only tries the two corners of each coordinate's interval and could step over the split points of a tree. The tree-aware search (procedure vi, 200 flows per family and seed) changes little: at $\epsilon = 0.2$ detection is 99.7\% for RF and 99.7\% for XGBoost, against 99.8\% and 99.7\% for SimBA on the same flows, and at $\epsilon = 0.5$ it is 99.3\% and 99.2\%. Its results are included in the worst-case robustness score used in Section V-D. To check whether coordinate-wise search simply misses joint evasions, we also solved the exact evasion problem for both tree ensembles as a mixed-integer program, following Kantchelian et al. [11], inside the same box as the other attacks: only the attacker-controllable features, the same budget, integer features kept integral and the min $\le$ mean $\le$ max orderings enforced (up to 200 flows per family and seed, 5 seeds, budgets 0.2, 0.5). A flow counts as certified when the program is infeasible, that is, when no point in the box falls below the detection threshold. At $\epsilon = 0.2$, certified detection is at least 99.8\% for RF and 100.0\% for XGBoost, and at most 99.9\% and 100.0\%. At $\epsilon = 0.5$, certified detection is at least 98.8\% for RF and 99.3\% for XGBoost, and at most 99.3\% and 99.7\%. Across all runs the solver found 55 verified evasions, and 49 flows were left undecided (time limit, or a solver solution that did not evade the real model). Within this threat model, then, the tree ensembles are robust in an exact sense, not merely against the searches we wrote, and their gap is not an artefact of a weak attack. We did not expect the trees to hold up this well, given that dedicated attacks evade tree ensembles [11]. A plausible explanation, which we have not verified, is that many of their splits use backward-direction and flag features that our threat model does not let the attacker touch. The exact search above shows that the robustness itself is real within the box; it does not tell us why.

\begin{table*}[t]\centering\caption{Detection rate at 1\% FPR under every perturbation procedure (mean $\pm$ 95\% CI over five seeds). $n$: attack flows per seed (all six families). Gradient attacks do not apply to tree ensembles (--); the tree-aware search applies only to them. Mixup: cross-family hybrids, no $\epsilon$.}\label{tab:pert}
\begin{tabular}{llrcccc}\toprule
Procedure & $\epsilon$ & $n$ & RF & XGBoost & 1D-CNN & FT-Transformer \\\midrule
Random sign & 0.05 & 14,642 & 1.000 $\pm$ 0.000 & 0.999 $\pm$ 0.000 & 0.999 $\pm$ 0.000 & 0.999 $\pm$ 0.000 \\
Random sign & 0.1 & 14,642 & 1.000 $\pm$ 0.000 & 0.999 $\pm$ 0.000 & 0.999 $\pm$ 0.000 & 0.999 $\pm$ 0.001 \\
Random sign & 0.2 & 14,642 & 1.000 $\pm$ 0.000 & 0.999 $\pm$ 0.000 & 0.999 $\pm$ 0.001 & 0.997 $\pm$ 0.003 \\
Random sign & 0.5 & 14,642 & 1.000 $\pm$ 0.000 & 0.999 $\pm$ 0.000 & 0.967 $\pm$ 0.020 & 0.988 $\pm$ 0.013 \\
FGSM (transfer) & 0.05 & 14,642 & 1.000 $\pm$ 0.000 & 1.000 $\pm$ 0.000 & 0.999 $\pm$ 0.000 & 0.999 $\pm$ 0.001 \\
FGSM (transfer) & 0.1 & 14,642 & 1.000 $\pm$ 0.000 & 0.999 $\pm$ 0.000 & 0.999 $\pm$ 0.001 & 0.998 $\pm$ 0.001 \\
FGSM (transfer) & 0.2 & 14,642 & 1.000 $\pm$ 0.000 & 0.999 $\pm$ 0.001 & 0.997 $\pm$ 0.002 & 0.996 $\pm$ 0.003 \\
FGSM (transfer) & 0.5 & 14,642 & 1.000 $\pm$ 0.000 & 0.999 $\pm$ 0.001 & 0.789 $\pm$ 0.115 & 0.973 $\pm$ 0.043 \\
PGD (transfer) & 0.05 & 14,642 & 1.000 $\pm$ 0.000 & 1.000 $\pm$ 0.000 & 0.999 $\pm$ 0.000 & 0.999 $\pm$ 0.001 \\
PGD (transfer) & 0.1 & 14,642 & 1.000 $\pm$ 0.000 & 0.999 $\pm$ 0.000 & 0.999 $\pm$ 0.001 & 0.998 $\pm$ 0.001 \\
PGD (transfer) & 0.2 & 14,642 & 1.000 $\pm$ 0.000 & 0.999 $\pm$ 0.001 & 0.997 $\pm$ 0.002 & 0.997 $\pm$ 0.002 \\
PGD (transfer) & 0.5 & 14,642 & 1.000 $\pm$ 0.000 & 0.999 $\pm$ 0.001 & 0.758 $\pm$ 0.110 & 0.974 $\pm$ 0.040 \\
PGD (white-box) & 0.05 & 14,642 & -- & -- & 0.998 $\pm$ 0.001 & 0.999 $\pm$ 0.001 \\
PGD (white-box) & 0.1 & 14,642 & -- & -- & 0.997 $\pm$ 0.001 & 0.998 $\pm$ 0.001 \\
PGD (white-box) & 0.2 & 14,642 & -- & -- & 0.970 $\pm$ 0.021 & 0.987 $\pm$ 0.012 \\
PGD (white-box) & 0.5 & 14,642 & -- & -- & 0.220 $\pm$ 0.094 & 0.824 $\pm$ 0.159 \\
SimBA & 0.05 & 2,253 & 0.999 $\pm$ 0.001 & 0.999 $\pm$ 0.001 & 0.997 $\pm$ 0.002 & 0.997 $\pm$ 0.003 \\
SimBA & 0.1 & 2,253 & 0.999 $\pm$ 0.001 & 0.998 $\pm$ 0.001 & 0.993 $\pm$ 0.003 & 0.994 $\pm$ 0.005 \\
SimBA & 0.2 & 2,253 & 0.998 $\pm$ 0.002 & 0.998 $\pm$ 0.002 & 0.909 $\pm$ 0.029 & 0.946 $\pm$ 0.054 \\
SimBA & 0.5 & 2,253 & 0.998 $\pm$ 0.002 & 0.993 $\pm$ 0.005 & 0.187 $\pm$ 0.127 & 0.667 $\pm$ 0.157 \\
Tree-aware search & 0.2 & 1,032 & 0.997 $\pm$ 0.002 & 0.997 $\pm$ 0.001 & -- & -- \\
Tree-aware search & 0.5 & 1,032 & 0.993 $\pm$ 0.009 & 0.992 $\pm$ 0.008 & -- & -- \\
Mixup & -- & 2,161 & 0.948 $\pm$ 0.010 & 0.934 $\pm$ 0.009 & 0.908 $\pm$ 0.050 & 0.706 $\pm$ 0.147 \\
\bottomrule\end{tabular}\end{table*}

The neural detectors are robust up to $\epsilon = 0.2$ (worst case 90.9\% for the 1D-CNN and 94.6\% for the FT-Transformer) but degrade sharply at $\epsilon = 0.5$: white-box PGD reduces detection to 22.0\% (1D-CNN) and 82.4\% (FT-Transformer), and SimBA to 18.7\% and 66.7\%. The 1D-CNN result is not driven by one bad seed; SimBA detection ranges from 7.1\% to 31.5\% across the five seeds. Transferred attacks from the surrogate are much weaker (75.8\% and 97.4\% at $\epsilon = 0.5$), and the random-sign control changes detection by at most a few points, so the degradation comes from the attack direction rather than from noise. Cross-family hybrids behave similarly: detection of mixup flows is 94.8\% (RF), 93.4\% (XGBoost), 90.8\% (1D-CNN) and 70.6\% (FT-Transformer).

\subsection{Detection of unseen attack families}

The LOAO picture is very different (Table~\ref{tab:loao}). Averaged over the six held-out families, detection at 1\% FPR is 76.2\% for RF, 40.2\% for XGBoost, 25.8\% for the 1D-CNN and 22.8\% for the FT-Transformer. Only DDoS is detected by every model when unseen, plausibly because its volumetric flow statistics resemble those of the DoS family still present in training. Unseen PortScan and BruteForce flows are detected at 10\% and 4\% by XGBoost and at most 7\% by the neural models; even RF, the best model in this regime, misses roughly half of them. The neural models also miss unseen DoS almost entirely (4\% and 4\%), whereas the tree ensembles detect it.

\begin{table}[t]\centering\caption{Detection rate at 1\% FPR on held-out (unseen) attack families, mean $\pm$ 95\% CI over seeds; $n$ is the number of test flows of the held-out family per seed. The Botnet and WebAttack rows rest on few flows and vary strongly between seeds; they are reported for completeness only.}\label{tab:loao}
\resizebox{\columnwidth}{!}{\begin{tabular}{lrcccc}\toprule
Held-out family & $n$ & RF & XGBoost & 1D-CNN & FT-Transformer \\\midrule
DoS & 5,000 & 0.988 $\pm$ 0.007 & 0.979 $\pm$ 0.010 & 0.043 $\pm$ 0.061 & 0.038 $\pm$ 0.023 \\
DDoS & 5,000 & 1.000 $\pm$ 0.000 & 1.000 $\pm$ 0.000 & 1.000 $\pm$ 0.000 & 1.000 $\pm$ 0.000 \\
PortScan & 2,309 & 0.481 $\pm$ 0.396 & 0.097 $\pm$ 0.009 & 0.058 $\pm$ 0.027 & 0.069 $\pm$ 0.015 \\
BruteForce & 2,080 & 0.573 $\pm$ 0.022 & 0.042 $\pm$ 0.104 & 0.000 $\pm$ 0.000 & 0.023 $\pm$ 0.060 \\
Botnet & 221 & 0.940 $\pm$ 0.030 & 0.206 $\pm$ 0.402 & 0.285 $\pm$ 0.482 & 0.176 $\pm$ 0.490 \\
WebAttack & 32 & 0.588 $\pm$ 0.428 & 0.087 $\pm$ 0.043 & 0.163 $\pm$ 0.154 & 0.062 $\pm$ 0.027 \\
\bottomrule\end{tabular}}\end{table}

The means hide what we found most telling: for several families, detection of an unseen family is close to all-or-nothing across seeds. The FT-Transformer detects unseen Botnet at 88.2\% in one seed and at 0.0\% in each of the other four; RF detects unseen PortScan at 80.8\%--85.1\% in two seeds and at 22.9\%--25.9\% in the other three. Whether a new family is caught can thus depend on which flows happened to be sampled for training, which is an uncomfortable property for a detector whose job is to catch the unexpected. It also means that the Botnet and WebAttack rows of Table~\ref{tab:loao}, with only 221 and 32 test flows per seed, have confidence intervals spanning most of the unit interval. We report them for completeness and draw no family-level conclusion from them; the per-model means over six families, and the paired comparison below, are far less sensitive to this.

Are the held-out families detectable at all at this operating point? The benign-only baselines suggest that, for most of them, not easily (Table~\ref{tab:anom}). The Isolation Forest detects on average 10.5\% of unseen-family flows and the autoencoder 4.6\%, although their AUROC is 0.88 and 0.90; only PortScan stands out (61.0\% and 17.9\%). This is less surprising than it looks: at a 1\% alert budget, a benign-only detector has to place an attack beyond the most unusual 1\% of benign flows, and benign traffic has a long tail of odd but harmless flows. The LOAO failures of the supervised detectors are therefore not simply an artefact of using discriminative models: anomaly detectors held to the same 1\% alert budget do worse on average, and RF does considerably better. The baselines are not robust either; under SimBA at $\epsilon = 0.5$ their already low detection falls to 0.5\% and 2.9\%.

\begin{table}[t]\centering\caption{Benign-only anomaly baselines on attack families they never saw: detection rate at 1\% FPR and AUROC, mean over five seeds.}\label{tab:anom}
\resizebox{\columnwidth}{!}{\begin{tabular}{lrcccc}\toprule
 & & \multicolumn{2}{c}{Isolation Forest} & \multicolumn{2}{c}{Autoencoder} \\
Family & $n$ & DR & AUROC & DR & AUROC \\\midrule
DoS & 5,000 & 0.003 & 0.884 & 0.047 & 0.855 \\
DDoS & 5,000 & 0.000 & 0.883 & 0.006 & 0.959 \\
PortScan & 2,309 & 0.610 & 0.973 & 0.179 & 0.974 \\
BruteForce & 2,080 & 0.000 & 0.867 & 0.000 & 0.886 \\
Botnet & 221 & 0.017 & 0.747 & 0.045 & 0.780 \\
WebAttack & 32 & 0.000 & 0.915 & 0.000 & 0.951 \\
\midrule Mean & & 0.105 & 0.878 & 0.046 & 0.901 \\
\bottomrule\end{tabular}}\end{table}

Would a detector built for unknown inputs do better? As an open-set baseline we retrained RF and XGBoost as multi-class models (benign plus one class per known family) and flagged a flow either when an attack class wins or when the winning benign class has low confidence (maximum-softmax-probability rejection, MSP), thresholded at the same 1\% FPR (5 seeds; Table~\ref{tab:openset}). At this operating point MSP gives the same detection rate as the simpler score $1 - P(\text{benign})$ in every seed, model and family: the two scores coincide whenever the benign class wins, and the 1\% FPR threshold is low enough that flows where an attack class wins are flagged under either score. Mean held-out-family detection is 67.0\% for RF and 52.7\% for XGBoost, against 76.2\% and 40.2\% for the binary detectors (-9.2 and +12.5 points). PortScan stays below 27\% and WebAttack below 28\% for both models. Open-set rejection of this simple kind therefore does not close the gap to the robustness scores. Stronger open-set methods exist [13], and we do not claim this baseline is the best possible; it shows that the gap is not specific to binary training.

\begin{table}[t]\centering\caption{Open-set baseline: multi-class RF and XGBoost with maximum-softmax-probability rejection (MSP), detection of the held-out family at 1\% FPR (mean over 5 seeds), next to the binary detectors of Table~\ref{tab:loao}. The score $1-P(\text{benign})$ gives identical detection rates at this threshold.}\label{tab:openset}
\begin{tabular}{lcccc}\toprule
 & \multicolumn{2}{c}{RF} & \multicolumn{2}{c}{XGBoost} \\
Held-out & Binary & MSP & Binary & MSP \\\midrule
DoS & 0.988 & 0.983 & 0.979 & 0.958 \\
DDoS & 1.000 & 1.000 & 1.000 & 1.000 \\
PortScan & 0.481 & 0.271 & 0.097 & 0.131 \\
BruteForce & 0.573 & 0.573 & 0.042 & 0.006 \\
Botnet & 0.940 & 0.910 & 0.206 & 0.899 \\
WebAttack & 0.588 & 0.281 & 0.087 & 0.169 \\
\midrule Mean & 0.762 & 0.670 & 0.402 & 0.527 \\
\bottomrule\end{tabular}\end{table}

\subsection{Robustness overstates, and only coarsely ranks, unseen-family detection}

For every model, worst-case robustness at $\epsilon \le 0.2$ exceeds mean LOAO detection by a large margin: RF 23.5 $\pm$ 7.5 points ($p < 10^{-3}$), XGBoost 59.5 $\pm$ 6.7 points ($p < 10^{-4}$), 1D-CNN 65.0 $\pm$ 12.8 points ($p < 10^{-3}$), FT-Transformer 71.8 $\pm$ 8.7 points ($p < 10^{-4}$) (paired t-tests over five seeds; Holm-adjusted across the four models, all $p \le 0.00096$; standardised effect sizes $d_z$ from 3.9 to 11.0). Every one of the five seed-level differences is positive for every model. With five seeds a per-model Wilcoxon test cannot fall below $p = 0.0625$, so we report the signed-rank test only over all 20 model-seed pairs, where all differences are positive and it reaches its smallest attainable value ($p = 1.9 \times 10^{-6}$). Used as an estimate of unseen-family detection in this setting, the robustness score would overstate it by 24 to 72 points (Table~\ref{tab:gapsens}).

\begin{table}[t]\centering\caption{Worst-case robustness ($\epsilon \le 0.2$, all procedures) against LOAO detection averaged three ways, and the resulting gap in percentage points (mean over five seeds; every model-seed gap is positive in all three cases).}\label{tab:gapsens}
\resizebox{\columnwidth}{!}{\begin{tabular}{lccccccc}\toprule
 & & \multicolumn{2}{c}{All six families} & \multicolumn{2}{c}{Four large families} & \multicolumn{2}{c}{Flow-weighted} \\
Model & Robustness & LOAO & Gap & LOAO & Gap & LOAO & Gap \\\midrule
RF & 0.997 & 0.762 & 23.5 & 0.760 & 23.6 & 0.851 & 14.5 \\
XGBoost & 0.997 & 0.402 & 59.5 & 0.530 & 46.7 & 0.700 & 29.6 \\
1D-CNN & 0.909 & 0.258 & 65.0 & 0.275 & 63.3 & 0.370 & 53.8 \\
FT-Transformer & 0.946 & 0.228 & 71.8 & 0.282 & 66.3 & 0.371 & 57.5 \\
\bottomrule\end{tabular}}\end{table}

Does robustness at least rank the detectors correctly? Only coarsely, and with little statistical support. The 20 model-seed pairs are not independent: they come from four models measured five times each. Over the four model means, Spearman's correlation is $\rho = 0.60$, and an exact permutation test over the 24 possible orderings gives $p = 0.21$; permuting whole models at the model-seed level gives $\rho = 0.69$ with $p = 0.04$. (A naive test that treats the 20 pairs as independent would report $p < 10^{-3}$, which overstates the evidence.) What both measures do agree on is that the tree ensembles sit above the neural models. The correlation is carried by that split between model families. Within the trees, XGBoost is as robust as RF (99.7\% vs. 99.7\%) but detects unseen families at roughly half RF's rate (40.2\% vs. 76.2\%); within the neural models the order reverses, with the FT-Transformer more robust and the 1D-CNN slightly better on unseen families. Robustness can tell a practitioner that, on this data and under our threat model, a tree ensemble is a safer choice than a small neural network. It cannot say which tree ensemble to deploy, and it says nothing reliable about how many new attacks the chosen model will catch.

The gap is not an artefact of the two small families (Table~\ref{tab:gapsens}). Averaging LOAO detection over the four large families only (DoS, DDoS, PortScan, BruteForce), the gap is 24 to 66 points; weighting families by their number of test flows, which favours the easily detected DoS and DDoS, it is 15 to 57 points. In both cases all 20 model-seed differences remain positive and every per-model paired t-test gives $p \le 0.003$.

The robustness score above pools all known-attack flows, so DoS and DDoS dominate it, whereas LOAO detection is an unweighted mean over families. Averaging robustness the same way, taking for each family the minimum over all attacks at $\epsilon \le 0.2$ and then the mean over the six families, gives 99.5\% (RF), 99.4\% (XGBoost), 84.7\% (1D-CNN), 88.4\% (FT-Transformer). The gap to mean LOAO detection is then 23.4 $\pm$ 5.9 points (RF), 59.2 $\pm$ 5.5 points (XGBoost), 58.9 $\pm$ 10.1 points (1D-CNN), 65.6 $\pm$ 11.7 points (FT-Transformer); the smallest single model-seed gap is 19.3 points and every per-model t-test gives $p \le 0.00092$. The neural detectors are the least robust on Botnet and WebAttack, which is where family averaging lowers their score, yet the gap stays large.

The gap also survives the temporal split: worst-case robustness minus mean LOAO detection is 28.9 points (RF), 61.0 points (XGBoost), 62.1 points (1D-CNN) there, against 23.6, 59.5, 65.0 for the random split on the same seeds. The temporal gap is positive in 15 of 15 model-seed pairs (paired t-tests per model, all $p \le 0.00051$). This comparison rests on 5 seeds and on the gradient and SimBA attacks only, so we read it as a robustness check of the main result rather than as a second estimate of its size. The FT-Transformer was not retrained under the temporal split, because only a CPU runtime was available.

\subsection{Feature consistency and repaired perturbations}

Feature-space perturbations can produce flows that no packet sequence could generate. We audited 8 derived-feature identities of CICFlowMeter (e.g., Flow Bytes/s = total bytes / duration), which hold for 100\% of clean test flows. After projection, 74\% to 93\% of perturbed or mixed flows violate an identity involving a modified feature, depending on variant and budget (averaged over identities), and already 74\% to 81\% at $\epsilon = 0.05$. Most perturbed flows in the main experiment are therefore not internally consistent. So we re-ran the perturbation experiments with all identities recomputed after every step (5 seeds; Table~\ref{tab:repair}). SimBA detection at $\epsilon = 0.5$ becomes 99.7\% for RF (99.8\% without repair), 99.4\% for XGBoost (99.3\% without repair), 20.7\% for 1D-CNN (18.7\% without repair), 65.6\% for FT-Transformer (66.7\% without repair). None of these changes is significant (paired t-tests over seeds, all $p \ge 0.19$). Repair, in other words, does not systematically weaken the attack. The repair run covers FGSM, PGD (transfer and white-box) and SimBA at every budget, not only SimBA at $\epsilon = 0.5$; the tree-aware search was not re-run with repair. Taking the worst case over FGSM, PGD (transfer and white-box) and SimBA at $\epsilon \le 0.2$ with identities enforced, robustness is 99.8\% for RF (99.8\% without repair), 99.8\% for XGBoost (99.8\% without repair), 94.3\% for 1D-CNN (90.9\% without repair), 93.2\% for FT-Transformer (94.6\% without repair), and the gap to mean LOAO detection becomes 23.6 $\pm$ 7.8 points (RF), 59.5 $\pm$ 6.7 points (XGBoost), 68.5 $\pm$ 11.9 points (1D-CNN), 70.4 $\pm$ 8.0 points (FT-Transformer), positive in every seed and model (paired t-tests, all $p \le 0.0011$). Using only consistent, repaired perturbations therefore leaves the conclusion unchanged. For the tree ensembles the unrepaired tree-aware result is within 0.1 points of SimBA, so leaving it out of the repaired comparison changes little.

\begin{table}[t]\centering\caption{Detection with and without enforcing the derived-feature identities (repair), mean $\pm$ 95\% CI over seeds. Worst case: minimum over FGSM, PGD (transfer, white-box) and SimBA at $\epsilon \le 0.2$. Gap: repaired worst case minus mean LOAO detection. $p$: paired t-tests.}\label{tab:repair}
\resizebox{\columnwidth}{!}{\begin{tabular}{lccccc}\toprule
 & \multicolumn{2}{c}{Worst case, $\epsilon \le 0.2$} & \multicolumn{2}{c}{SimBA, $\epsilon = 0.5$} & \\
Model & No repair & Repair & No repair & Repair & Gap (repair) \\\midrule
RF & 0.998 & 0.998 & 0.998 & 0.997 ($p$=0.59) & 0.236 $\pm$ 0.078 \\
XGBoost & 0.998 & 0.998 & 0.993 & 0.994 ($p$=0.19) & 0.595 $\pm$ 0.067 \\
1D-CNN & 0.909 & 0.943 & 0.187 & 0.207 ($p$=0.23) & 0.685 $\pm$ 0.119 \\
FT-Transformer & 0.946 & 0.932 & 0.667 & 0.656 ($p$=0.56) & 0.704 $\pm$ 0.080 \\
\bottomrule\end{tabular}}\end{table}

\subsection{Threshold-free metrics overstate unseen-family performance}

In LOAO, mean AUROC remains high for the tree ensembles (0.963 for RF and 0.966 for XGBoost) while their detection rates at 1\% FPR are 76.2\% and 40.2\%. For example, unseen PortScan obtains an AUROC of 0.949 with XGBoost but only 10\% detection at the operating point, and the anomaly baselines show the same pattern. Unseen attacks are ranked above most benign traffic but not above its tail, so an operator working with a realistic alert budget would miss them. LOAO results should therefore be reported at a fixed false-positive rate rather than only as AUROC.

\subsection{A second dataset: CSE-CIC-IDS2018}

Every result so far comes from one dataset. We therefore repeated the main experiment, unchanged, on the improved CSE-CIC-IDS2018 release of the same authors [9]. The release has about 10 GB of compressed flows, more than our 12 GB of memory can hold, so we drew a uniform random sample within each family, keeping every flow of a family with the same probability (at most 1.5 million benign flows and 250,000 flows per attack family; 1,440,158 unique flows after the same cleaning). The held-out families are DoS, DDoS, PortScan, BruteForce, Botnet, WebAttack (3 seeds; RF, XGBoost, 1D-CNN; the FT-Transformer was not run because only a CPU runtime was available; Table~\ref{tab:cic18}). Known-attack detection is 100.0\% (RF), 100.0\% (XGBoost), 99.9\% (1D-CNN), worst-case robustness at $\epsilon \le 0.2$ is 99.8\%, 99.1\%, 96.3\%, and mean LOAO detection is 83.3\%, 71.0\%, 28.5\%. The gap is 16.5 points (RF; 3 of 3 seeds positive, $p = 0.0087$), 28.1 points (XGBoost; 3 of 3 seeds positive, $p = 0.001$), 67.9 points (1D-CNN; 3 of 3 seeds positive, $p = 0.0035$). For RF the gap is smaller than on CIC-IDS2017 (16.5 vs. 23.5 points). For XGBoost the gap is smaller than on CIC-IDS2017 (28.1 vs. 59.5 points). Here the tree ensembles detect most unseen families almost completely and miss mainly PortScan and WebAttack, whereas the 1D-CNN again misses most of them. The direction of the result therefore repeats on a second, independently captured dataset: robustness stays high for every model, while unseen-family detection falls short of it in every seed, by a margin that depends on the model and the dataset.

\begin{table}[t]\centering\caption{CSE-CIC-IDS2018 (uniform per-family sample; 3 seeds): known-attack detection, worst-case robustness ($\epsilon \le 0.2$; FGSM, PGD, SimBA), LOAO detection per held-out family and on average, and the gap in points.}\label{tab:cic18}
\resizebox{\columnwidth}{!}{\begin{tabular}{lcccccccccc}\toprule
Model & Known & Robust. & DoS & DDoS & PortScan & BruteForce & Botnet & WebAttack & LOAO & Gap \\\midrule
RF & 1.000 & 0.998 & 0.999 & 1.000 & 0.445 & 1.000 & 1.000 & 0.557 & 0.833 & 16.5 \\
XGBoost & 1.000 & 0.991 & 0.993 & 0.943 & 0.179 & 1.000 & 0.999 & 0.145 & 0.710 & 28.1 \\
1D-CNN & 0.999 & 0.963 & 0.060 & 0.406 & 0.112 & 0.073 & 0.670 & 0.388 & 0.285 & 67.9 \\
\bottomrule\end{tabular}}\end{table}

\subsection{Our view and recommendations}

We would keep reporting robustness to perturbations; it measures something real, namely how stable a decision boundary is around attacks the model already knows. What we would stop doing is citing it as evidence of zero-day capability. Holding out an attack family costs a few extra training runs. In our experiments the FT-Transformer kept 94.6\% worst-case detection at $\epsilon \le 0.2$ yet detected only 22.8\% of held-out-family flows on average; a robustness report alone would not have revealed that. Concretely:

\begin{itemize}
\item Report robustness to perturbations and detection of held-out attack families as separate results; neither substitutes for the other.
\item Report LOAO detection at a fixed, realistic false-positive rate in addition to AUROC, with per-seed values, since unseen-family detection can vary strongly between seeds.
\item Evaluate tree ensembles with query-based attacks rather than gradient attacks they cannot receive, and audit perturbed flows for internal consistency.
\item Use corrected datasets and remove duplicate and label-conflicting flows before splitting.
\end{itemize}

\section{Limitations}
Our perturbations are feature-space. Enforcing the eight derived-feature identities (Section V-E) removes the most obvious inconsistencies, but it still does not guarantee that a real packet sequence exists for each perturbed flow [21]; problem-space attack generation would be the stronger test. The attacker-controllable set also includes flow-level aggregates, such as packet-length statistics and Flow Bytes/s, that depend partly on the victim's responses; an attacker can move them only through its own packets. Restricting the attacker to forward-direction features alone would weaken the attack, which could only raise the robustness scores and widen the gap, so this choice is conservative with respect to our conclusion. The label-conflict step uses labels from the whole dataset, but it removes only 28 flows (14 BENIGN, 14 PortScan), fewer than 0.01\% of the data; the other 384,746 removed flows are exact duplicates of a flow with the same label. Its effect on any result is therefore negligible. Splits in the main experiment are random at the flow level after deduplication. A temporal split on 5 seeds leaves the gap in place (Section V-D), but it omits the FT-Transformer, and a session-level split remains to be done. A random split does not leak the held-out family into LOAO training, since that family is absent from training altogether. On the attack side, our black-box attacker sees continuous scores and may query freely, which is more than a deployed NIDS usually reveals, so the threat model errs towards a strong attacker; the remaining risk is on the search side: the exact MILP search (Section V-B) removes the search-side doubt for the tree ensembles, but it covers 5 seeds and up to 200 flows per family, and the gradient and SimBA attacks on the neural detectors remain heuristic, so their robustness may be overstated. Results come from two corrected datasets, with the second covering fewer seeds and models, and from binary detectors. WebAttack and Botnet test sets are small (32 and 221 flows per seed) and their LOAO results vary strongly between seeds, so we do not draw family-level conclusions from them. Model capacity and training budgets are moderate, and stronger neural models may behave differently. Finally, we did not evaluate defences: adversarial training could raise the robustness side of the comparison, and domain-adaptation or open-set methods could raise the LOAO side. Whether either narrows the gap is an open and, in our view, the most useful next question.

\section{Conclusion}
On corrected CIC-IDS2017, four detectors with near-perfect accuracy on known attacks remain largely robust to constrained feature-space perturbations, yet miss 24\% to 77\% of flows from attack families absent from training. On this benchmark, robustness overstates unseen-family detection by 24 to 72 percentage points for every model and agrees with it only on the coarse ranking of trees over neural networks. A second dataset, CSE-CIC-IDS2018, shows the same direction with a gap of 16 to 68 points, smaller for the tree ensembles than on CIC-IDS2017. The two benign-only anomaly detectors we tested did not close the gap at a 1\% alert budget. Within the scope tested here (two corrected datasets, the second with three detectors and three seeds, and a feature-space threat model), perturbations of known attacks measure a different property from zero-day generalization, and synthetic perturbation sets should not be offered as evidence that a NIDS detects new attacks. We see this as one instance of a broader problem in security machine learning: benchmark numbers that are easy to produce get used to answer questions they were never designed to answer. Held-out-family evaluation at a fixed false-positive rate is cheap and reproducible, and we hope it becomes routine. The next steps we consider most useful are testing whether adversarial training or open-set and domain-adaptation methods narrow the gap, and repeating the protocol with more datasets and with problem-space attack generation.

\section{Appendix: Attacker-controllable features}
The 42 attacker-controllable features (of 82) are: Flow Duration; Total Fwd Packet; Total Length of Fwd Packet; Fwd Packet Length Max; Fwd Packet Length Min; Fwd Packet Length Mean; Fwd Packet Length Std; Flow Bytes/s; Flow Packets/s; Flow IAT Mean; Flow IAT Std; Flow IAT Max; Flow IAT Min; Fwd IAT Total; Fwd IAT Mean; Fwd IAT Std; Fwd IAT Max; Fwd IAT Min; Fwd Header Length; Fwd Packets/s; Packet Length Min; Packet Length Max; Packet Length Mean; Packet Length Std; Packet Length Variance; Average Packet Size; Fwd Segment Size Avg; Fwd Bytes/Bulk Avg; Fwd Packet/Bulk Avg; Fwd Bulk Rate Avg; Subflow Fwd Packets; Subflow Fwd Bytes; Fwd Act Data Pkts; Active Mean; Active Std; Active Max; Active Min; Idle Mean; Idle Std; Idle Max; Idle Min; Total TCP Flow Time. The remaining 40 features (backward-direction statistics, TCP flag counts, protocol, initial window sizes, the down/up ratio and the forward minimum segment size) are never modified.

\section*{Acknowledgment}
The author used an AI-based research assistant to help draft and edit portions of this manuscript and to write and run the experiment code under the direction of the author. The author designed the study, checked the code, data, analyses and conclusions, and takes full responsibility for the content.

\end{document}